\documentclass[12pt]{article}
\usepackage{amstext,amsmath,amssymb}

\newtheorem{theorem}{Theorem}
\newtheorem{proposition}[theorem]{Proposition}
\newtheorem{definition}[theorem]{Definition}
\newtheorem{corollary}[theorem]{Corollary}

\def\qed{\hspace*{\fill} $\Box$\par\medskip}

\newcommand{\newfontobj}[2]{
  \newcommand{#1}[1]{
    \expandafter\def\csname##1\endcsname{\ensuremath{#2{##1}}}}}

\newfontobj{\class}{\mathbf}
\newfontobj{\defi}{\mathrm}
\class{co}\class{BB}
\class{P} \class{FP} \class{UP} \class{FewP} \class{NP}
\class{SAT} \class{coNP} \class{PSPACE} \class{EXP} \class{QBF}
\class{TFNP}
\class{TF} \class{TQBF}
\class{F}
\class{PH}
\defi{bad}

\newcommand{\tuple}[1]{{\langle{#1}\rangle}}

\begin{document}

\title{Compression Complexity}

\author{Stephen Fenner\\University of South Carolina \and Lance Fortnow\\Georgia Institute of Technology}

	\maketitle

\begin{abstract}
	The Kolmogorov complexity of $x$, denoted $C(x)$, is the length of the shortest program that generates $x$. For such a simple definition, Kolmogorov complexity has a rich and deep theory, as well as applications to a wide variety of topics including learning theory, complexity lower bounds and SAT algorithms.
	
	Kolmogorov complexity typically focuses on decompression, going from the compressed program to the original string. This paper develops a dual notion of compression, the mapping from a string to its compressed version. Typical lossless compression algorithms such as Lempel-Ziv or Huffman Encoding always produce a string that will decompress to the original. We define a general compression concept based on this observation.
	
	For every $m$, we exhibit a single compression algorithm $q$ of length about $m$ which for $n$ and strings $x$ of length $n\geq m$, the output of $q$ will have length within $n-m+O(1)$ bits of $C(x)$. We also show this bound is tight in a strong way, for every $n \geq m$ there is an $x$ of length $n$ with $C(x)\approx m$ such that no compression program of size slightly less than $m$ can compress $x$ at all.
	
	We also consider a polynomial time-bounded version of compression complexity and show that similar results for this version would rule out cryptographic one-way functions. 
	
\end{abstract}

\section{Introduction}


Kolmogorov complexity has a rich history, with many applications to areas such as computability, machine learning, number theory, and computational complexity.  The book of Li \&\ Vit\'anyi~\cite{LV} gives an excellent background on the subject.

Kolmogorov complexity measures the information content inherent in a string $x$ by the length of the shortest program that computably produces $x$; this length is denoted $C(x)$. We think of the program $p$ as a compressed version of $x$ and the algorithm producing $x$ a decompression procedure. Kolmogorov complexity focuses on decompression procedures. In this paper we turn our attention to the task of producing $p$ given $x$, i.e., compression.  

How hard is it to compute the shortest program $p$ from $x$? It's equivalent to the halting problem since even the set $R$ of random strings ($x$ for which $C(x)\geq |x|$) is hard for the halting problem~\cite{Kolmogorov65,Chaitin,Solomonoff1,Solomonoff2}.

If we are given the value $C(x)$ then we can compute a program $p$ of that length by dovetailing through all programs of that length until we find one the produces $x$. While this method works on a single string, it does not work in general and won't even halt if we are given an underestimate of the length.

Consider the lossless compression algorithms in use today. Among the general-purpose compression algorithms are Huffman coding, run-length encoding, Burrows-Wheeler, and Lempel-Ziv (and variants thereof) \cite{wiki:compression}. All of these algorithms have a common property, for any input $x$ they will produce an output that will decompress back to $x$.

To this end we define compression functions as having this single property. A \emph{compression function} is a computable function $q$ such that $U(q(x))=x$ for all strings $x$, for some fixed universal Turing machine $U$. The trivial compression function $q(x)$ simply outputs the program ``Print $x$''.

A compression algorithm may benefit from having information embedded in it; for example, a Huffman encoding tree (of letters, digrams, or even trigrams) to Huffman compress English text.  Therefore, the size of a compression algorithm itself may improve its performance---at least on some strings.

We investigate a basic trade-off between the size of a general-purpose compression algorithm and how well it compresses.

Our main result shows roughly that for any value of $m$, there is a compression function $q$ of size roughly $m$ that will fully compress any string of Kolmogorov complexity at most $m$. We need only $m$ bits for $q$ even though there are exponentially many (in $m$) strings that $q$ can fully compress. 

We also show an essentially tight strong inverse: roughly, that for any $n$ and $m$ with $m < n$ there is a single string $x$ of length $n$ and Kolmogorov complexity $m$ such that every compression function $q$ of size less than $m$ fails to compress $x$ at all. 

We further consider polynomial-time bounded versions of compression complexity whose results depend on unproven assumptions in computational complexity. If $\P=\NP$, then one can have perfect efficient polynomial-time compression functions; on the other hand, even a small compression would break one-way functions. 

Finally we explore some future directions for research into compression complexity.

\subsection{Main Results}
\label{mainresultssec}

Our first theorem says that for any $m$ there is a compression function of size about $m$ that optimally compresses all strings $x$ with $C(x) \le m$ and does not significantly expand other strings.

In what follows, $k$ is a sufficiently large fixed constant independent of $m$ or $x$.  See the next section for detailed definitions.

\begin{theorem}\label{upperthm}
	For every $m$, there exists a compression function $q$ with $|q| \leq m+k$ such that, for all strings $x$,
	\begin{enumerate}
		\item $|q(x)| = C(x)$ if $C(x) \le m$, and
		\item $|q(x)| \leq |x| + k$ otherwise.
	\end{enumerate}
\end{theorem}

\begin{corollary}\label{uppercor}
	For every $m$, there exists a compression function $q$ with $|q| \leq m +k$ such that, for all $x$ with $|x|\geq m$, 
	\[|q(x)|-C(x) \leq |x|-m+k. \]
\end{corollary}

The next theorem implies that Theorem~\ref{upperthm} and is corollary are essentially tight.  It says for any $n\ge m$ that if a compression function $q$ for length $n$ (i.e., one that only needs to work on strings of length $n$) is significantly shorter than $m$, then it behaves poorly on at least one string: there is an $x$ of length $n$---independent of $q$---such that $C(x)$ is about $m$ but $q$ does not compress $x$ at all.

\begin{theorem}\label{lowerthm}
	For all $m,n$ with $0 \leq m\leq n$ there exists an $x$ such that
	\begin{enumerate}
		\item $|x| = n$,
		\item \label{c-item} $C(x) \leq m + k \log n$, 
		\item \label{compitem} For all compression functions $q$ for length $n$, with $|q| \leq m - k\log n$,
		$|q(x)| \geq n$, and 
		\item \label{diffitem} For all compression functions $q$ for length $n$, with $|q| \leq m - k\log n$,
		$|q(x)|-C(x) \geq n - m - k\log n$.
	\end{enumerate}
\end{theorem}
Note that (\ref{diffitem}) follows from (\ref{c-item}) and (\ref{compitem}).  Compare (\ref{diffitem}) with the result of Corollary~\ref{uppercor}.

We prove Theorems~\ref{upperthm} and~\ref{lowerthm} in Section~\ref{proofsec}. In Section~\ref{timesec} we give definitions and results for time-bounded compression.

\section{Preliminaries}

Our notation and basic definitions are standard.  Here we briefly outline the basic definitions of Kolmogorov complexity.  For an in-depth treatment, see the standard textbook on the subject by Li \& Vit\'anyi~\cite{LV}.

We assume all strings are over the binary alphabet $\{0,1\}$.  We define a standard pairing function $\tuple{\cdot,\cdot}$ injectively mapping pairs of strings to strings as
\[ \tuple{x,y} := 1^{|x|}0xy \]
for all strings $x$ and $y$.

Fix a universal machine $U$ suitable for defining Kolmogorov complexity.  It suffices to choose $U$ such that, for any Turing machine $M$, there exists a string $p$ such that $M(x) = U(\tuple{p,x})$ for any string $x$ such that $M(x)$ is defined.  Abusing notation, we also write $p(x)$ for $U(\tuple{p,x})$ and call $p$ a ``program.''

Define $C(x)$ as the length of the shortest $p$ such that $U(p)=x$.

Both our main results freely refer to a constant $k$.  We can (and do) choose $k$ large enough (depending only on $U$) such that our results hold.

\begin{definition}\rm
Let $n$ be a natural number.  A \emph{compression function for length $n$} is a program $q$ such that $U(q(z))=z$ for all strings $z$ of length $n$.  A \emph{compression function} is a program $q$ that is a compression function for all lengths.
\end{definition}
Note that any compression function $q$ is total, and for all $x$, \ $q(x)\geq C(x)$.

Let $\BB(m)$ (``Busy Beaver of $m$'') be the maximum time for $U(p)$ to halt over all halting programs $p$ of length at most $m$. Let $p_m$ be the lexicographically least program of length at most $m$ that achieves $\BB(m)$ running time.


\section{Proof of the Main Results}
\label{proofsec}

In this section we give proofs of Theorems~\ref{upperthm} and~\ref{lowerthm}.

\medskip

{\bf Proof of Theorem~\ref{upperthm}}: 

Given $m$, we define $q(z)$ as follows:
\begin{itemize}
\item Let $t$ be the number of steps used by $U(p_m)$ before halting. Note that $t=\BB(m)$. 
\item Look for the lexicographically shortest program $p$ of length at most $|z|$ such that $U(p)=z$ within $t$ steps.
\begin{itemize}
\item If $p$ is found, then output $p$.
\item Otherwise, output ``Print $z$.''
\end{itemize}
\end{itemize}
We hardwire the value of $p_m$ into the code for $q$ (e.g., let $q = \tuple{r,p_m}$ for some program $r$ independent of $m$), so $|q|\leq m+k$.

Fix a string $x$.

{\bf Case 1: }$C(x)\leq m$.

Let $p$ be the lexicographically first program of length $C(x)$ such that $U(p)=x$. By the definition of $\BB(m)$, $U(p)$ will halt in at most $t$ steps.
So we'll have $q(x)=p$ and $|q(x)|=C(x)$. 

{\bf Case 2: }$C(x) > m$.

Either $q(x)$ outputs ``Print $x$'' or a program of length at most $n$. Either way $|q(x)|\leq n+k$.
%
\qed

\bigskip

{\bf Proof of Theorem~\ref{lowerthm}}: 

Let $A_s^\ell$ be the set of strings $y$ of length $\ell$ such that there is no program $p$ with $|p|<\ell$ such that $U(p)$ outputs $y$ within $s$ steps.  One can compute a canonical index for $A_s^\ell$ given $s$ and $\ell$ as input.  $A_s^\ell$ contains the random strings of length $\ell$, so in particularly $A_s^\ell$ is never empty. If $s\geq \BB(\ell)$, then $A_s^\ell$ is exactly the set of random strings of length $\ell$.

Given $m \le n$, let $t=\BB(m)$ and $x$ be the lexicographically first string in $A^n_t$.

Note that $C(x)\leq m+k\log n$,
since we can describe $x$ by $p_m$, $m$, and $n$, using $p_m$ to find $t$.

Suppose there is a compression function $q$ for length $n$ with $|q| \leq m - k\log n$ such that $|q(x)| < n$.

Let $z$ be the lexicographically first random string of length $m$. We show how to use $q$, $n$, and $m$ to find $z$. This will yield a contradiction to the fact that $z$ is random.

Let $t'$ be the maximum over all $y$, $|y|=n$, of the number of steps required for $U$ to halt on input $q(y)$. Here we use the fact that $U(q(y))=y$ for all $y$ of length $n$.

Since $|q(x)|<n$ and $U(q(x))=x$, by the definition of $x$ the number of steps $\hat{t}$ required for $U$ to halt on input $q(x)$ must be greater than $t=\BB(m)$. So we have $t'\geq \hat{t}> t=\BB(m)$. We can't necessarily compute $t$ or $\hat{t}$ from just $q$, $n$, and $m$, but we can compute $t'$.

Now compute $A_{t'}^m$, which will be exactly the random strings of length $m$, and $z$ will be the lexicographically least string in that set.

\qed

\section{Time-Bounded Compression Complexity}
\label{timesec}

The proofs in Section~\ref{proofsec} create machines that run in time based on the busy-beaver function, which grows faster than any computable function. Practical compression and decompression algorithms need to be far more efficient. We explore a polynomial time-bounded version of compression complexity in this section, though, not too surprisingly, the results we get will depend on open questions in computational complexity.

Time bounds can play into both the compression and decompression procedures.

\begin{definition}\rm
	An \emph{$(f,g)$-time bounded compression function} is a function $q$ such that for all strings $x$,
	\begin{enumerate}
		\item $q(x)$ halts within $f(|x|)$ steps.
		\item $U(q(x))$ halts and outputs $x$ within $g(|x|)$ steps.
	\end{enumerate}
\end{definition}

Let $C^t(x)$ be the length of the shortest program $p$ such that $U(p)$ outputs $x$ within $t(|x|)$ steps. For every $(f,g)$-compression function $q$, \ $|q(x)|\geq C^g(x)\geq C(x)$. 

Typically we consider $f$ and $g$ as polynomials. For a fixed polynomial $p$ we can easily compute the smallest $p$-time bounded program for $x$ in $\FP^\NP$. As a corollary we get
\begin{proposition}
	If $\P = \NP$ then for every polynomial $p$ there is a polynomial $p'$ and a $(p',p)$-compression function $q$ such that $|q(x)|=C^p(x)$. 
\end{proposition}
So in particular, to show that we don't have perfect efficient compression would require settling the $\P$ versus $\NP$ problem.

On the other hand, we show that if we can have short compression programs that more than trivially compress all strings that have significant compressions, \textit{\`a la} Theorem~\ref{upperthm} and Corollary~\ref{uppercor}, then we can break one-way functions.

First we need to define a family of compression functions.

\begin{definition}\rm
	A \emph{family of $(f,g)$-time bounded compression functions} is an enumeration of programs $q_0,q_1,\ldots$ such that for all $n$ and all strings $x$ of length $n$,
	\begin{enumerate}	
		\item $q_n(x)$ halts within $f(n)$ steps.
		\item $U(q_n(x))$ halts and outputs $x$ within $g(n)$ steps.
	\end{enumerate}
We say a compression family has \emph{polynomial-size} if for some constant $c$, $|q_n|\leq n^c$ for all $n\ge 2$.
A compression family is \emph{uniform} if there is a polynomial-time algorithm $Q$ such that $Q(1^n)=q_n$ for all $n$.
\end{definition}

\begin{theorem}
\label{factorthm}
	Fix a constant $\delta$ with $0<\delta<1$. Suppose that for any polynomial $p'$ there is a polynomial $p$ and a polynomial-size family of $(p,p')$-compression functions $q_0,q_1,\ldots$ such that for all $n$ and $x\in\Sigma^n$ with $C^{p'}(x)\leq n^\delta$, we have $|q_n(x)|< n$.  Then one-way functions do not exist relative to polynomial-size circuits.
	
	In particular, we could factor numbers on average with polynomial-size circuits. Under the additional assumption that the family is uniform we can factor numbers on average with a polynomial-time algorithm.
\end{theorem}

 
{\bf Proof:} Let's assume we have a one-way function $f$. H\aa{}stad, Impagliazzo, Levin and Luby~\cite{HILL} show how to convert this one-way function into a polynomial-time pseudorandom generator $G:\Sigma^{n^\epsilon}\rightarrow\Sigma^n$ (for any fixed $\epsilon>0$) so that no polynomial-size circuit can distinguish the output of $G$ on a random seed from a truly uniformly chosen string of length $n$.

Choose an $\epsilon$ so that $0<\epsilon<\delta$. Pick $p'(n)$ larger than the running time of $G$ and let $q_0,q_1,\ldots$ be the family of $(p,p')$-compression functions given in the assumptions of Theorem~\ref{factorthm}. 

Consider the following test $T(x)$ that can be expressed as a polynomial-sized circuit: Output $0$ if $|q_n(x)|<n$, and output $1$ otherwise.

Buhrman, Jiang, Li and {Vit\'{a}nyi}~\cite{BJLV} show that a constant fraction of the strings of length $n$ have $C(x)\geq n$. Since $|q_n(x)|\geq C(x)$, if we choose a string $x$ at random, then $T(x)$ will output 1 with probability at least some constant $\alpha>0$.
 
Suppose $x$ is the output of $G(r)$ for some $r$. We can describe $x$ by the code for $G$ and $r$, or $n^\epsilon+O(1)\leq n^\delta$ bits for sufficiently large $n$. Since the running time of $G$ is less than $p'$, we have $|C^{p'}(x)|\leq n^\delta$. By the assumptions of Theorem~\ref{factorthm} we have $|q(x)|<n$ and $T(x)=0$.

$T(x)$ will output $1$ with probability at least $\alpha$ when $x$ is chosen at random and with probability $0$ when $x$ is the output of $G$ on a randomly chosen seed. This contradicts the fact that $G$ is a pseudorandom generator and $f$ is a one-way function.

In particular, the function that maps two primes to their product is not one-way, so we can factor randomly chosen numbers using polynomial-size circuits.

If the family of compression functions is uniform, then $T(x)$ above can be expressed as a polynomial-time algorithm.  H\aa{}stad, Impagliazzo, Levin and Luby~\cite{HILL} show how to take any one-way function against polynomial-time algorithms and convert it to a pseudorandom generator that has no uniform tests. Putting this together there must be a polynomial-time algorithm that factors randomly chosen numbers. \qed

\section{Future Directions}
\label{opensec}

This work is just the start of compression complexity. 

Is there a compression analogue of conditional Kolmogorov complexity $C(x|y)$? The results of Section~\ref{mainresultssec} should go through if we allow $q$ to have access to $y$, but it is less clear what happens if $q$ does not have access to $y$.

Symmetry of Information shows that for any strings $x$ and $y$, \ $C(x,y)\leq C(x)+C(y|x)+O(\log(|x|+|y|))$. Suppose we consider compression functions $q(x,y)$ to produce programs $p_1$ and $p_2$ such that $U(p_1)=x$ and $U(\tuple{p_2,x})=y$. How short can we get $|p_1|+|p_2|$ compared to $C(x,y)$?

There are several other notions of time-bounded Kolmogorov complexity (see~\cite{LV}) such as distinguishing complexity, where we need only distinguish a string $x$ from other strings. Do the results of Section~\ref{timesec} still apply?
 
One could also consider lossy compression, perhaps building on Kolmo\-gorov complexity with errors~\cite{FLV}.
 
\subsection*{Acknowledgments}

We would like to thank Eric Allender, Saurabh Sawlani, and Jason Teutsch for helpful discussions.

\bibliographystyle{alpha}
\bibliography{compression}

\end{document}